\documentstyle[prl,aps,psfig,multicol]{revtex}
\begin{document}
\draft
\title{Classical versus Quantum Effects in the $B=0$ Conducting Phase in 
Two Dimensions}
\author{S.~V.~Kravchenko\cite{svk}, D.~Simonian\cite{ds}, K.~Mertes, and 
M.~P.~Sarachik} 
\address{Physics Department, City College of the City University of New York, New York, New York 10031}
\author{T.~M.~Klapwijk}
\address{Department of Applied Physics and Materials Science Centre, 
University of Groningen,\\ 9747 AG Groningen, The Netherlands}
\date{\today}
\maketitle
\begin{abstract}
In the dilute two-dimensional electron system in silicon, we show that the temperature below which Shubnikov-de~Haas oscillations become apparent is approximately the same as the temperature below which an exponential decrease in resistance is seen in $B=0$, suggesting that the anomalous behavior in zero field is observed only when the system is in a degenerate (quantum) state.  The temperature dependence of the resistance is found to be qualitatively similar in $B=0$ and at integer Landau level filling factors.
\end{abstract}
\pacs{PACS numbers: 71.30.+h, 73.40.Qv, 73.40.Hm}
\begin{multicols}{2}

The anomalous transport behavior observed
\cite{kravchenko,simonian97a,hanein98,simmons98a} in a number of different dilute two-dimensional (2D) electron and hole systems has been the focus of considerable attention since its discovery several years ago.  Although many theoretical suggestions have been advanced 
\cite{dobrosavljevic97,he98,lyandageller98,castellani98,phillips98,si98,altshuler98,klapwijk98,thakur98,dassarma98}, no generally accepted explanation has emerged.  The anomalous behavior is observed only in high-mobility ``clean'' 2D systems and is characterized by two
main features: (i)~for an electron (or hole) density, $n_s$, exceeding some (material-dependent) critical value, $n_c$, the resistivity drops by a factor of up to ten as the temperature $T$ is decreased below some temperature, $T_M$; and (ii)~the decrease in resistance at $T<T_M$ is completely suppressed by a magnetic field, while the transport is independent of field at temperatures above $T_M$.

Theoretical work in the early 1980's\cite{finkelshtein84} pointed to the possible existence of a metallic phase in 2D when interactions between particles are strong, a conducting state that would be suppressed by a magnetic field.  However, the mechanism suggested by these 
theories\cite{castellani98,finkelshtein84} predicts an onset for the metallic phase at temperatures that are lower than those found experimentally: in most systems, $T_M\sim\frac{1}{3}T_F$, while in
Refs.~\cite{castellani98,finkelshtein84}, the condition
$T<<\hbar/k_B\tau<<T_F$ was assumed (here $T_F=E_F/k_B$ is the Fermi temperature, $E_F$ is the Fermi energy, and $\tau$ is the lifetime of the electron).  This discrepancy was addressed in Ref.\cite{si98} where it was shown that the metal-insulator transition should be evident at temperatures of the order of the Fermi temperature.  Non-Fermi liquid states \cite{dobrosavljevic97}, including superconductivity \cite{phillips98}, have also been suggested to account for the experimental observations.  Finally, several ``classical'' models have been proposed \cite{altshuler98,klapwijk98,dassarma98} in which the scattering of the electrons in charged traps plays a crucial role.

If the ``metallic'' behavior (as well as its suppression by a magnetic field) is observed when the electron system is non-degenerate, all the anomalous effects most likely have a classical origin, and quantum explanations could be ruled out.  There is a simple way to check experimentally whether a 2D system is in a degenerate (``quantum'') or non-degenerate (``classical'') state: it is clearly degenerate if strong Shubnikov-de~Haas oscillations are observed in a perpendicular magnetic field~\cite{degenerate}.  In this paper, we compare the temperature $T_M$, below which the sharp decrease in resistance is observed in $B=0$, with the temperature which marks the onset of 
Shubnikov-de~Haas oscillations {\em at the same low density of electrons}.  We find that these temperatures are quite close, establishing that the anomalous drop in resistance in the absence of a magnetic field occurs when the electrons condense into a degenerate ``quantum'' system~\cite{caveat}.  We show further that the temperature dependence of the resistance at $B=0$ and at the integer Landau level filling factors are quite similar.

The silicon metal-oxide-semiconductor field-effect tran\-sistors
(MOSFET's) used in these studies were especially prepared for measurements at low electron densities and low temperatures.  The electron density in the contact regions in these samples can be varied independently of the electron density $n_s$ in the main channel and, therefore, it is possible to maintain low resistance of the contact regions even when the main part of the sample is at very low electron density.  This is achieved by making very narrow gaps in the gate metallization.  All samples had electron mobilities at $T=4.2$~K between $22,000$ and $25,000$~cm$^2$/Vs.

Figure~\ref{1} shows the diagonal resistivity, $\rho$, of a dilute 2D electron system in silicon as a function of perpendicular magnetic field at several temperatures.  At the highest temperature, 1.5~K (dotted line), the resistance decreases slightly as the magnetic field is increased up to $B=3.5$~T, and then quickly grows.  Note that the weak magnetoresistance at small fields remains essentially the same in size as the temperature is reduced from 1.5~K to 0.5~K, indicating that it is not associated with the suppression of weak localization.  At $T=1.5$~K there are no minima in $\rho(B)$ at magnetic fields corresponding to integer filling factors of Landau levels.  This suggests either that the Landau levels are broader than the splitting between them, or that the electron system is not degenerate at this relatively high temperature.  Indeed, at an electron density of
$0.74\times10^{11}$~cm$^{-2}$, the Fermi temperature is about 5~K, only a factor of three larger than the temperature~\cite{energy}.  Weak oscillations of $\rho(B)$ begin to develop at $T=1$~K, and grow as the temperature is reduced.  The deep minima at temperatures of 500~mK and below at magnetic fields close to $B=1.6$ and 3.1~T correspond to Landau level filling factors $\nu=2$ and 1.  The electrons are clearly degenerate at these temperatures.

Except for the resistivity at the highest temperature, the $\rho(B)$ curves cross at a single point at $B=3.5$~T, signaling a quantum Hall (QHE) to insulator transition (see, {\it e.g.}, Ref.~\cite{shahar95}).  The value of $\rho$ at the crossing point in different 2D systems is generally close to $h/e^2$ \cite{shahar95}.  It is interesting to note that in our system it is $\approx5.5\times10^4$~Ohm, close to the value of the resistivity at the $B=0$ metal-insulator transition.

The upper curve of Fig.~\ref{2}~(a) shows the resistivity as a function of temperature in zero magnetic field for the same low electron density as Fig.~\ref{1}.  The resistivity decreases below about $1$ K, or roughly the temperature below which the Shubnikov-deHaas oscillations in Fig.~\ref{1} become evident.  This implies that the onset of the anomalous resistance drop occurs at approximately the temperature below which the electron system is beginning to condense into a degenerate quantum state.

Qualitatively similar behavior has been found in low-density 
p-GaAs/AlGaAs heterostructures, where the energy scale and $T_M$ are several times smaller\cite{simmons98a}.  There are no oscillations in $\rho(B)$ at $T=500$~mK; they become noticeable at $T=333$~mK and develop into quantum Hall minima at lower temperatures.  For the same hole density, the decrease in resistance at $B=0$ occurs below approximately 300~mK, a temperature quite comparable with the temperature below which the quantum Hall minima develop.  We note also that the initial magnetoresistance is temperature-independent in 
p-GaAs/AlGaAs, as it is in our silicon MOSFET's.

We now compare the behavior of the resistivity in zero field with its temperature dependence in a magnetic field corresponding to a quantum Hall minimum.  Figures~\ref{2}~(a) and \ref{2}~(b) show $\rho(T)$ at $B=0$ and at $\nu=2$ for two different electron densities.  In both cases, $\rho(T)$ grows as the temperature is decreased, reaches a maximum around $T_M$ ($T_M\approx3$~K at
$n_s=0.93\times10^{11}$~cm$^{-2}$ and $\approx1$~K at $n_s=0.74\times10^{11}$~cm$^{-2}$), and rapidly drops at lower temperatures.  The resistance at $\nu=1$ (not shown) has the same shape  although the decrease is deeper and $T_M$ is higher.  In contrast, the temperature-dependence is very different at intermediate values of magnetic field, as shown by the dotted curve of Fig.~\ref{2}~(a).

There is a strong similarity between the curves shown in Fig.~\ref{2} in zero field and for a field where $\nu=2$.  While the behavior at $\nu=2$ reflects the formation of a quantum Hall state, there is no consensus at this point concerning the origin of the resistance decrease in the absence of a magnetic field.  Various explanations have been proposed, including a purely classical model \cite{altshuler98} which attributes the resistance drop to the (temperature-dependent) neutralization of positively charged traps in the silicon oxide layer.  Given the uncanny similarity between $\rho(T)$ at $B=0$ and at $\nu=2$, it would be remarkable if the curve at $\nu=2$ were of quantum origin while the curve in $B=0$ were due to purely classical effects involving the unintended and uncontrolled presence of charged traps.  The fact that the drop in resistance in zero magnetic field appears {\it only} when the electron system is degenerate also argues against the model proposed in Ref.~\cite{altshuler98}, which gives equally strong ``metallic'' temperature dependence in degenerate and non-degenerate regimes.  It is consistent with the model proposed in Ref.~\cite{dassarma98}, which predicts ``insulating'' temperature-dependence for a non-degenerate case and ``metallic'' dependence for a degenerate case (see, however, comment \cite{kravchenko98a}).

One could argue that the similarity between $\rho(T)$ at $B=0$ and $\nu=2$ is accidental and due simply to the fact that the resistance in both cases decreases from a moderately high, almost $B$-independent, ``classical'' value above $T_M$ to a smaller value in the degenerate (``quantum'') state as $T\rightarrow0$.  The latter is known to be zero at $\nu=2$ but is unknown for the case $B=0$.  On the other hand, there may be a deeper connection between the $B=0$ and QHE states.  In fact, it has been shown recently \cite{hanein98b} that the critical point for a finite-field QHE-insulator transition evolves continuously into the critical point for the zero-field metal-insulator transition as the electron density is increased.

To summarize, we have shown that (i) the temperature $T_M$ below which the anomalous decrease in the resistivity is observed in the 
two-dimensional electron system in silicon is approximately the same as the temperature below which Shubnikov-de~Haas oscillations become manifest, indicating that the electrons are degenerate; (ii)~the resistivity displays qualitatively similar temperature dependences in zero magnetic field and in fields corresponding to the integer QHE minima.  These findings both suggest that the anomalous sharp decrease of the resistance in the absence of a magnetic field is an intrinsic property of the electron system and is observed only when the system is in the quantum state.

We are grateful to S.~Bakker and R.~Heemskerk for their contributions in developing and preparation of the MOSFET's used in this work.  This work was supported by the US Department of Energy under Grant 
No.~DE-FG02-84ER45153.  Partial support was also provided by NSF Grant 
No.~DMR-9803440.

\begin{figure}
\caption{Diagonal resistivity as a function of perpendicular\\ magnetic
field at five temperatures:  1.5~K, 1.0~K, 0.75~K, 0.5~K,\\ and 0.24~K. 
The electron density  $n_s=0.74\times10^{11}$~cm$^{-2}$.}
\label{1}
\end{figure}
\begin{figure}
\caption{Temperature dependences of the diagonal resistivity\\ at $B=0$
and at Landau level filling factor $\nu=2$ for:\\ 
(a)~$n_s=0.74\times10^{11}$~cm$^{-2}$; and 
(b)~$n_s=0.93\times10^{11}$~cm$^{-2}$.\\  The dotted line in (a) shows the
resistivity at $B=1.12$~Tesla.}
\label{2}
\end{figure}
\end{multicols}

\begin{thebibliography}{10}
\bibitem{degenerate} The observation of Shubnikov-de~Haas oscillations also requires that the splitting between Landau levels be larger than their width.
\bibitem{caveat} We note that this condensation does not by itself produce any change in resistivity, since the Drude formula gives the same value of conductivity for classical (Boltzmann) and quantum (Fermi) statistics, nor does it explain the anomalous effects observed in response to magnetic field\cite{simonian97a} in the ``quantum'' state.
\bibitem{energy} We have assumed that the effective mass is not increased by electron-electron interactions which would lead to even smaller $E_F$.  We also note that $E_F$ changes in a magnetic field and increases up to $2E_F$ when all the electrons are fully 
spin-polarized.
\end{thebibliography}
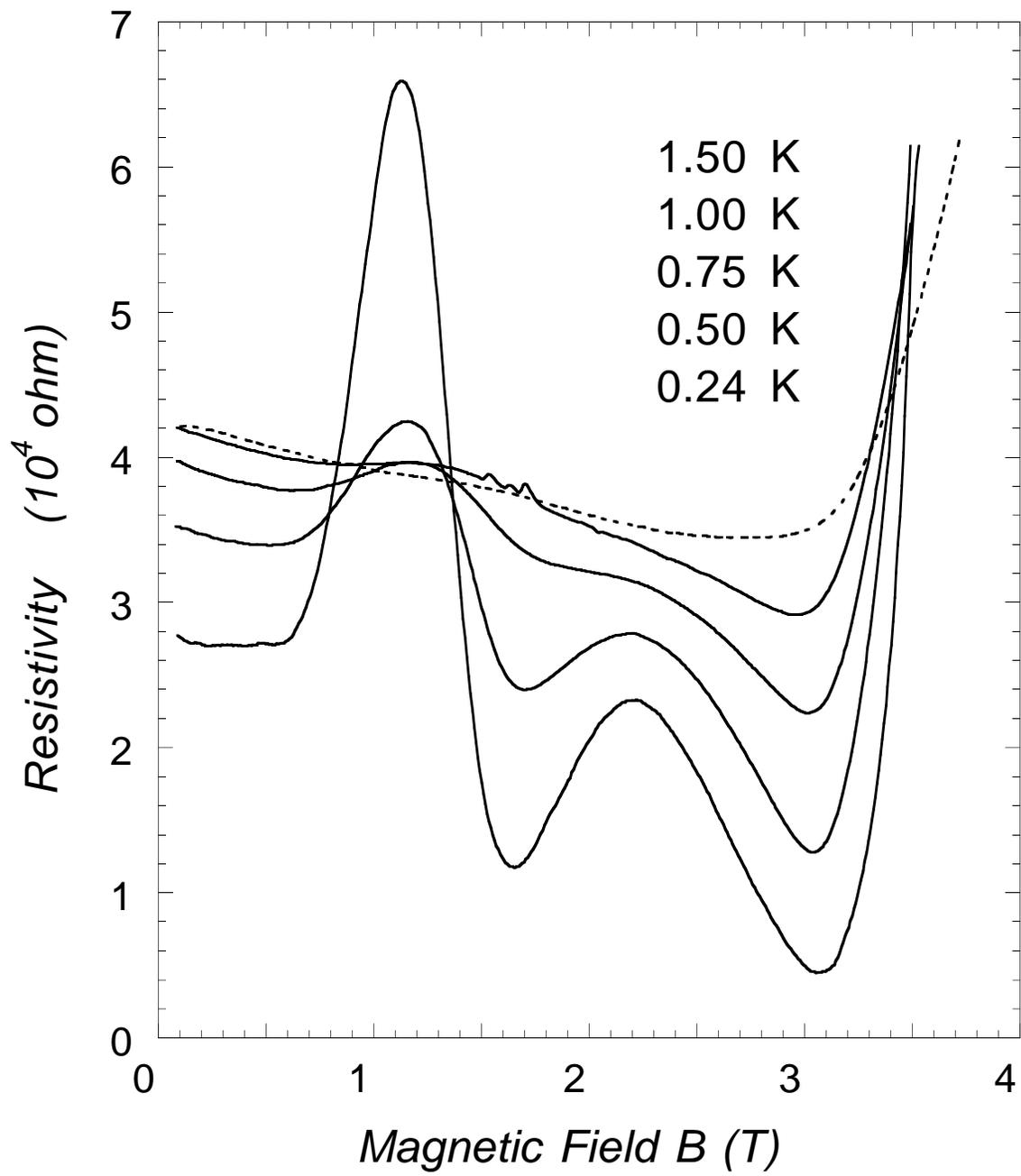
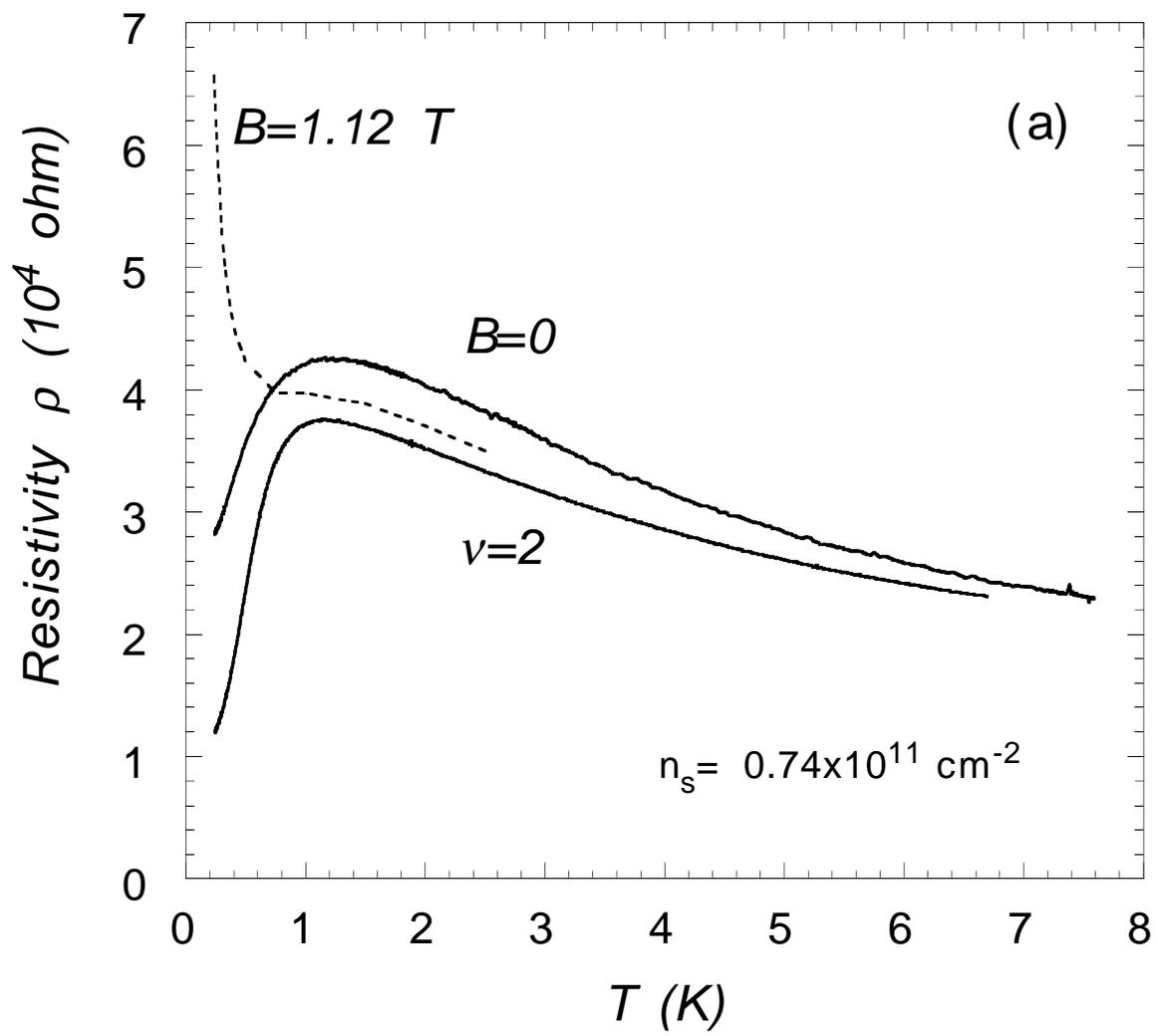
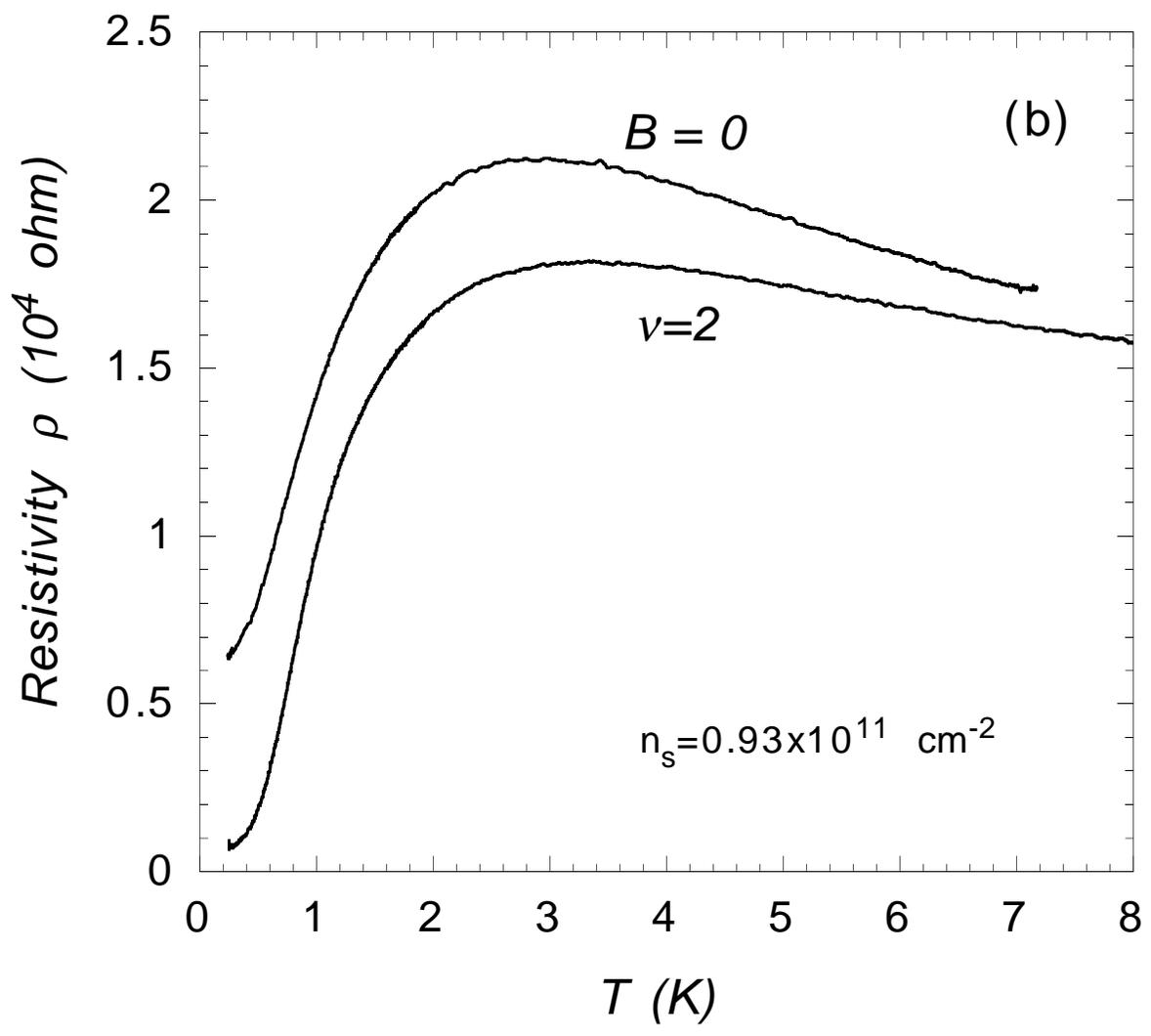
\end{document}